\documentclass[proceedings]{JHEP37}
\usepackage{eurosym}
\usepackage{amssymb}
\usepackage{amsfonts}
\usepackage{amsmath}
\usepackage{epsfig}

\newbox\mybox

\newcommand\fverb{\setbox\mybox=\hbox\bgroup\verb}
\newcommand\fverbdo{\egroup\medskip\noindent\fbox{\unhbox\mybox}\ }
\newcommand\fverbit{\egroup\item[\fbox{\unhbox\mybox}]}
\conference{A new non-Hermitian E2-quasi-exactly solvable model}
\abstract{We construct a previously unknown $E_2$-quasi-exactly solvable non-Hermitian model 
whose eigenfunctions involve weakly orthogonal polynomials obeying three-term recurrence relations that
factorize beyond the quantization level. The model becomes Hermitian when one of its two parameters is fixed to a specific value. We analyze the double scaling limit of this model leading to the complex Mathieu equation. The norms, Stieltjes measures and moment functionals are evaluated for some concrete values of one of the two 
parameters.}

\title{A new non-Hermitian E2-quasi-exactly solvable model}
\author{Andreas Fring \\
Department of Mathematics, City University London,\\
Northampton Square, London EC1V 0HB, UK\\
E-mail: a.fring@city.ac.uk}

\input{tcilatex}
\begin{document}

In \cite{E2Fring} we introduced $E_{2}$-quasi-exactly solvable models in
analogy to the notion of $sl_{2}(\mathbb{C})$-quasi-exactly solvability
originally proposed by Turbiner \cite{Turbiner00,Tur0}. The different
setting is motivated mathematically by the fact that solutions for $E_{2}$%
-quasi-exactly solvable models do not belong to the general class of
hypergeometric functions which emerge as solutions from an $sl_{2}(\mathbb{C}%
)$-setting. The physical motivation results from the current interest in
extending the study of solvable models \cite%
{KhareM,BijanMQR,BenMou,BijanQR,E2Fring} to non-Hermitian quantum mechanical
systems \cite{Urubu,Bender:1998ke,Benderrev,Alirev}. The $E_{2}$%
-quasi-exactly solvable models are especially interesting in optical
settings \cite%
{Muss,MatMakris,Guo,OPMidya,MatHugh,MatHughEva,DFM,DFM2,MatLongo} where the
fact is exploited that the Helmholtz equation results as a reduction from
the Schr\"{o}dinger equation. Solvable models are rare exceptions in the
study of quantum mechanical systems and the model presented here should be
added to that list.

The starting point for the construction of $E_{2}$-quasi-exactly systems
consists of expressing the Hamiltonian operator $\mathcal{H}$ of the model
in terms of the $E_{2}$-basis operators $u$, $v$ and $J$ obeying the
commutation relations 
\begin{equation}
\left[ u,J\right] =iv,\qquad \left[ v,J\right] =-iu,\qquad \left[ u,v\right]
=0.
\end{equation}%
instead of the standard $sl_{2}(\mathbb{C})$-generators. We now use the
particular realization \cite{Isham} 
\begin{equation}
J:=-i\partial _{\theta },\quad u:=\sin \theta ,\quad v:=\cos \theta ,
\label{Ish}
\end{equation}%
and demand a specific anti-linear symmetry \cite{EW}, $\mathcal{PT}_{3}:$ $%
J\rightarrow J$, $u\rightarrow v$, $v\rightarrow u$, $i\rightarrow -i$, as
defined in \cite{DFM}, which for (\ref{Ish}) becomes $\mathcal{PT}%
_{3}:\theta \rightarrow \pi /2-\theta $, $i\rightarrow -i$. The operators in
(\ref{Ish}) act on the $\mathcal{PT}_{3}$-invariant vector spaces over $%
\mathbb{R}$ 
\begin{eqnarray}
V_{n}^{s}(\phi _{0}) &:&=\limfunc{span}\left\{ \left. \phi _{0}\left[ \sin
(2\theta ),i\sin (4\theta ),\ldots ,i^{n+1}\sin (2n\theta )\right]
\right\vert \theta \in \mathbb{R},\mathcal{PT}_{3}(\phi _{0})=\phi _{0}\in
L\right\} ,~~~~~~  \label{q1} \\
V_{n}^{c}(\phi _{0}) &:&=\limfunc{span}\left\{ \left. \phi _{0}\left[
1,i\cos (2\theta ),\ldots ,i^{n}\cos (2n\theta )\right] \right\vert \theta
\in \mathbb{R},\mathcal{PT}_{3}(\phi _{0})=\phi _{0}\in L\right\} .
\label{q2}
\end{eqnarray}%
Taking the groundstate eigenfunction to be $\phi _{0}^{c}=e^{i\kappa \cos
2\theta }$ with $\kappa \in \mathbb{R}$ we identified in \cite{E2Fring} the
following actions of combinations of the basis operators $%
J:V_{n}^{s,c}\left( \phi _{0}^{c}\right) \mapsto V_{n+1}^{c,s}\left( \phi
_{0}^{c}\right) $, $uv:V_{n}^{s,c}\left( \phi _{0}^{c}\right) \mapsto
V_{n+1}^{c,s}\left( \phi _{0}^{c}\right) $ and $i(u^{2}-v^{2}):V_{n}^{s,c}%
\left( \phi _{0}^{c}\right) \mapsto V_{n+1}^{s,c}\left( \phi _{0}^{c}\right) 
$. Quasi-exact solvability is achieved if we are able to impose suitable
constraints such that $\mathcal{H}_{N}:V_{\tilde{n}}^{s,c}\left( \phi
_{0}^{c}\right) \mapsto V_{\tilde{n}}^{s,c}\left( \phi _{0}^{c}\right) $ for
specific values $\tilde{n}$. It remains a challenge to construct new models
of $E_{2}$-type that satisfy the latter condition.

We introduce here the $\mathcal{PT}_{3}$-symmetric Hamiltonian%
\begin{equation}
\mathcal{H}_{N}=J^{2}+\zeta uvJ+2i\zeta N(u^{2}-v^{2}),\qquad \zeta ,N\in 
\mathbb{R},  \label{newH}
\end{equation}%
and demonstrate explicitly that it is $E_{2}$-quasi-exactly solvable. We
notice from the above that $\mathcal{H}_{N}:V_{n}^{s,c}\left( \phi
_{0}^{c}\right) \mapsto V_{n+2}^{s,c}\left( \phi _{0}^{c}\right) \oplus
\zeta V_{n+2}^{s,c}\left( \phi _{0}^{c}\right) \oplus V_{n+1}^{s,c}\left(
\phi _{0}^{c}\right) $, such that it appears to be possible to reduce the
order of the target space by imposing two additional constraints. In
general, the Hamiltonian $\mathcal{H}_{N}$ is non-Hermitian except for $%
N=1/4 $, with free $\zeta \in \mathbb{R}$, which we deduce from the fact
that $\mathcal{H}_{N}^{\dagger }=\mathcal{H}_{1/2-N}$. A further interesting
feature of this Hamiltonian is that it reduces to the complex Mathieu
Hamiltonian in the double scaling limit $\lim_{N\rightarrow \infty ,\zeta
\rightarrow 0}\mathcal{H}_{N}=\mathcal{H}_{\text{Mat}%
}=J^{2}+2ig(u^{2}-v^{2}) $ for $g:=N\zeta <\infty $ similarly as the
Hamiltonian discussed in \cite{E2Fring,BijanQR}.

According to (\ref{q1}) and (\ref{q2}) we make the Ansatz 
\begin{eqnarray}
\psi _{N}^{c}(\theta ) &=&e^{\frac{i}{4}\zeta \cos (2\theta
)}\sum_{n=0}^{\infty }i^{n}\frac{P_{n}}{\zeta ^{n}N(1+2N)_{n-1}}\cos
(2n\theta ),  \label{psis} \\
\psi _{N}^{s}(\theta ) &=&ie^{\frac{i}{4}\zeta \cos (2\theta
)}\sum_{n=1}^{\infty }i^{n}\frac{Q_{n}}{\zeta ^{n}N(1+2N)_{n-1}}\sin
(2n\theta ),\text{\quad }  \label{psic}
\end{eqnarray}%
for our eigenfunctions with $P_{n}$, $Q_{n}$ being polynomials to be
determined and $(a)_{n}:=\Gamma \left( a+n\right) /\Gamma \left( a\right) $
denoting the Pochhammer symbol. The denominators have been extracted in such
a way that $P_{n}$ and $Q_{n}$ become $n$-th order polynomials in $E$ when $%
\psi _{N}$ is substituted into Schr\"{o}dinger equation $\mathcal{H}_{N}\psi
_{N}=E\psi _{N}$. In this way we obtain the three-term recurrence relations%
\begin{eqnarray}
P_{1} &=&EP_{0},  \label{r1} \\
P_{n+1} &=&2(E-4n^{2})P_{n}+\zeta ^{2}\left[ 4N^{2}-2N-n(n-1)\right]
P_{n-1},\quad \text{for }n=1,2,3,\ldots  \label{r2} \\
Q_{2} &=&2(E-4)Q_{1},  \label{r3} \\
Q_{n+1} &=&2(E-4n^{2})Q_{n}+\zeta ^{2}\left[ 4N^{2}-2N-n(n-1)\right]
Q_{n-1},\quad \text{for }n=2,3,4,\ldots  \label{r4}
\end{eqnarray}%
These equations may be solved in general as outlined in \cite{E2Fring}.
Taking $P_{0}=1$ we obtain for the lowest orders 
\begin{eqnarray}
P_{1} &=&E, \\
P_{2} &=&2E^{2}-8E+2\zeta ^{2}N(2N-1),  \notag \\
P_{3} &=&4E^{3}-80E^{2}+E\left( 2\zeta ^{2}\left( 6N^{2}-3N-1\right)
+256\right) +64\zeta ^{2}(1-2N)N,  \notag \\
P_{4} &=&8E^{4}-448E^{3}+E^{2}\left[ 16\zeta ^{2}(N-1)(2N+1)+6272\right]
-192E\left[ \zeta ^{2}\left( 6N^{2}-3N-1\right) +96\right]  \notag \\
&&+4\zeta ^{2}N(2N-1)\left[ \zeta ^{2}(N+1)(2N-3)+1152\right] .  \notag
\end{eqnarray}%
Likewise with $Q_{1}=1$ we compute 
\begin{eqnarray}
Q_{2} &=&2E-8, \\
Q_{3} &=&4E^{2}-80E+2\zeta ^{2}(N-1)(2N+1)+256,  \notag \\
Q_{4} &=&8E^{3}-448E^{2}+8E\left[ \zeta ^{2}(\allowbreak 2N^{2}-N-2)+784%
\right] -32\left[ \zeta ^{2}(10N^{2}-5N-6)+576\right] ,  \notag \\
Q_{5} &=&16E^{4}-1920E^{3}+8E^{2}\left[ \zeta ^{2}\left( 6N^{2}-3N-10\right)
+8736\right]  \notag \\
&&-32E\left[ \zeta ^{2}(94N^{2}-47N-106)+26240\right]  \notag \\
&&+4\left[ \zeta ^{4}(4N^{4}-4N^{3}-13N^{2}+7N+6)+128\zeta
^{2}(82N^{2}-41N-54)+589824\right] .  \notag
\end{eqnarray}%
We observe the typical feature for quasi-exactly solvable systems that the
three term relation can be reset to a two-term relation at a certain level.
This is due to the fact that in (\ref{r2}) and (\ref{r4}) the last term
vanishes when $n=2N$. Thus when taking $N$ to be a half-integer, $N\in 
\mathbb{N}/2$, we find the typical factorization%
\begin{equation}
P_{2N+n}=R_{n}P_{2N}\qquad \text{and\qquad }Q_{2N+n}=R_{n}Q_{2N}.
\end{equation}%
The first solutions for the factor $R_{n}$ are 
\begin{eqnarray}
R_{1} &=&2E-32N^{2}, \\
R_{2} &=&4E^{2}-16E\left( 8N^{2}+4N+1\right) +4N\left[ 64N(2N+1)^{2}-\zeta
^{2}\right] .
\end{eqnarray}%
Thus our polynomials $P_{n}$ and $Q_{n}$ possess the standard properties of
Bender-Dunne polynomials \cite{Bender:1995rh}.

Let us now determine the energy eigenvalues $E_{2N}$ from the conditions $%
P_{2N}(E)=0$ and $Q_{2N}(E)=0$ for the lowest values of $N$. For the
solutions related to (\ref{psis}) we compute 
\begin{eqnarray}
E_{1}^{c} &=&0, \\
E_{2}^{c,\pm } &=&2\pm \sqrt{4-\zeta ^{2}}, \\
E_{3}^{c,\ell } &=&\frac{20}{3}+\frac{2\hat{\Omega}}{3}e^{\frac{i\pi \ell }{3%
}}-\frac{2}{3}\left( 3\zeta ^{2}-52\right) e^{-\frac{i\pi \ell }{3}}\hat{%
\Omega}^{-1},\qquad ~~  \label{E5}
\end{eqnarray}%
with $\ell =0,\pm 2$ and $\hat{\Omega}:=\left[ 280+36\zeta ^{2}+3^{3/2}\sqrt{%
\zeta ^{6}-4\zeta ^{4}+1648\zeta ^{2}-2304}\right] ^{1/3}$ etc. and for the
solutions related to (\ref{psic}) we obtain 
\begin{eqnarray}
E_{2}^{s} &=&4, \\
E_{3}^{s,\pm } &=&10\pm \sqrt{36-\zeta ^{2}}, \\
E_{4}^{s,\ell } &=&\frac{56}{3}+\frac{2}{3}e^{\frac{i\pi \ell }{3}}\Omega -%
\frac{2}{3}e^{-\frac{i\pi \ell }{3}}\Omega ^{-1}\left( 3\zeta
^{2}-196\right) ,\qquad ~~  \label{E7}
\end{eqnarray}%
with $\ell =0,\pm 2$ and $\Omega :=\left[ 1144+36\zeta ^{2}+3^{3/2}\sqrt{%
\zeta ^{6}-148\zeta ^{4}+15856\zeta ^{2}-230400}\right] ^{1/3}$ etc.

The exceptional points are computed from the real zeros of the discriminants 
$\Delta _{2N}^{c}$ and $\Delta _{2N}^{s}$ for the polynomials $P_{2N}(E)$
and $Q_{2N}(E)$, respectively, 
\begin{eqnarray}
\Delta _{2}^{c} &=&\zeta ^{2}-4,  \label{d1} \\
\Delta _{3}^{s} &=&\zeta ^{2}-36,  \notag \\
\Delta _{3}^{c} &=&\zeta ^{6}-4\zeta ^{4}+1648\zeta ^{2}-2304,  \notag \\
\Delta _{4}^{s} &=&\zeta ^{6}-148\zeta ^{4}+15856\zeta ^{2}-230400,  \notag
\\
\Delta _{4}^{c} &=&\zeta ^{12}+8\zeta ^{10}+6160\zeta ^{8}-2119680\zeta
^{6}+4128768\zeta ^{4}-749850624\zeta ^{2}+530841600,  \notag \\
\Delta _{5}^{s} &=&\zeta ^{12}-376\zeta ^{10}+2^{4}7041\zeta
^{8}-2^{11}11925\zeta ^{6}+2^{13}207675\zeta ^{4}-2^{12}19579725\zeta
^{2}+2^{18}2480625.  \notag  \label{d6}
\end{eqnarray}%
We have suppressed here overall constant factors that do not contribute to
the values of the zeros. Our numerical for the solutions of these equations
multiplied by $N$ are presented in table 1.

\begin{table}[h]
\begin{center}
\begin{tabular}{c||c|c|c|c|c|}
$N$ & $\zeta _{0}^{c}N$ & $\zeta _{0}^{s}N$ & $\zeta _{0}^{c}N$ & $\zeta
_{0}^{s}N$ & $\zeta _{0}^{c}N$ \\ \hline\hline
$1$ & $2.00000$ &  &  &  &  \\ \hline
$3/2$ & $1.77556$ & $9.00000$ &  &  &  \\ \hline
$2$ & $1.68457$ & $8.21937$ & $21.0567$ &  &  \\ \hline
$5/2$ & $1.63564$ & $7.8691$ & $19.4554$ & $38.2224$ &  \\ \hline
$3$ & $1.6047$ & $7.6688$ & $18.6864$ & $35.5683$ & $60.535$ \\ \hline
& $\vdots $ & $\vdots $ & $\vdots $ & $\vdots $ & $\vdots $ \\ \hline
$\infty $ & $1.46877$ & $6.92895$ & $16.4711$ & $30.0967$ & $47.806$%
\end{tabular}%
\end{center}
\caption{Values of $\protect\zeta _{0}N$ computed from the positive real
zeros $\protect\zeta _{0}$ of the discriminant polynomials $\Delta
_{2N}^{c,s}$ approaching the critical values of the complex Mathieu
equation. }
\label{T1}
\end{table}

We observe that in the double scaling limit the critical values for the
Mathieu equation seem to be approached from above, albeit from further away
than in \cite{E2Fring}. As also noted in \cite{E2Fring} a much better
convergence can be obtained when instead of computing successively the
exceptional points for each level one takes the limit directly for the
three-term recurrence relation. Thus carrying out the limit $N\rightarrow
\infty $, $\zeta \rightarrow 0$\ with $g:=N\zeta <\infty $ on (\ref{r1})-(%
\ref{r4}) with the additional assumption that the coefficient functions
remain finite, i.e. $\lim_{N\rightarrow \infty ,\zeta \rightarrow
0}P_{n}=:P_{n}^{M}$ and $\lim_{N\rightarrow \infty ,\zeta \rightarrow
0}Q_{n}=:Q_{n}^{M}$ we obtain%
\begin{eqnarray}
P_{1}^{M} &=&EP_{0}^{M}  \label{ei1} \\
-2gP_{n-1}^{M}+4n^{2}P_{n}^{M}+\frac{1}{2}P_{n+1}^{M} &=&EP_{n}^{M},
\label{ei2} \\
-2gQ_{n-1}^{M}+4n^{2}Q_{n}^{M}+\frac{1}{2}P_{n+1}^{M} &=&EQ_{n}^{M}.
\label{ei3}
\end{eqnarray}%
The recurrence relations may be viewed as two eigenvalue equations for the
infinite matrices $\Xi $ and $\Theta $ with entries%
\begin{eqnarray}
\Xi _{i,j} &=&4i^{2}\delta _{i,j}+\frac{1}{2}\delta _{j,i+1}-2g^{2}\delta
_{i,j+1},\quad \quad \qquad \qquad ~\text{for }i,j\in \mathbb{N}, \\
\Theta _{i,j} &=&4i^{2}\delta _{i,j}+\frac{1}{2}\delta _{j,i+1}-2g^{2}\delta
_{i,j+1}+\frac{1}{2}\delta _{i,0}\delta _{j,1},\quad \text{for }i,j\in 
\mathbb{N}_{0},
\end{eqnarray}%
acting on the vectors $(Q_{1}^{M},Q_{2}^{M},Q_{3}^{M},\ldots )$ and $%
(P_{0}^{M},P_{1}^{M},P_{1}^{M},\ldots )$, respectively. The real zeros $%
g_{0} $ of the discriminants $\Delta ^{\Xi }(g)$ and $\Delta ^{\Theta }(g)$
of the characteristic polynomials $\det (\Xi ^{\ell }-E\mathbb{I})$ and $%
\det (\Theta ^{\ell }-E\mathbb{I})$ for the truncated matrices correspond to
the exceptional points. The matrices differ from those reported in \cite%
{E2Fring}, denoted here $\Xi ^{\lbrack 1]}$ and $\Theta ^{\lbrack 1]}$,
obtained from the double scaling limit analyzed therein. However, as the
difference is simply $2g\Xi _{i,i+1}=\Xi _{i,i+1}^{[1]}$, $\Xi _{i+1,i}=\Xi
_{i+1,i}^{[1]}2g$ the products $\Xi _{i,i+1}\Xi _{i+1,i}=\Xi
_{i,i+1}^{[1]}\Xi _{i+1,i}^{[1]}$ remain invariant and thus the matrices
posses the same eigenvalues. We may argue in a similar way for $\Theta $.
Thus the critical values are identical to those reported in tables 2 and 3
of \cite{E2Fring}.

Using the linear functional $\mathcal{L}$ \cite{Favard,Finkel} acting on
arbitrary polynomials $p$ as%
\begin{equation}
\mathcal{L}(p)=\int\nolimits_{-\infty }^{\infty }p(E)\omega (E)dE,\qquad 
\mathcal{L}(1)=1,  \label{L}
\end{equation}%
the standard norm $N_{n}^{\Phi }$ for the orthogonal polynomials $\Phi
_{n}(E)$ is defined via%
\begin{equation}
\mathcal{L}(\Phi _{n}\Phi _{m})=N_{n}^{\Phi }\delta _{nm}.  \label{ortho}
\end{equation}%
The normalization in (\ref{L}) implies $N_{0}^{P}=1$ and $N_{1}^{Q}=1$. The
three-term recurrence relations (\ref{r1})-(\ref{r4}) together with (\ref%
{ortho}) lead to%
\begin{eqnarray}
\quad \!\!\!N_{n}^{P} &=&\mathcal{L}(P_{n}^{2})=\mathcal{L}%
(EP_{n-1}P_{n})=\prod\limits_{k=1}^{n}b_{k}=\frac{\zeta ^{2n}}{2}%
(1-2N)_{n}(2N)_{n},~~\quad n=2,3,... \\
\!\!\!N_{n}^{Q} &=&\mathcal{L}(Q_{n}^{2})=\mathcal{L}(EQ_{n-1}Q_{n})=\prod%
\limits_{k=2}^{n}b_{k}=\frac{\zeta ^{2n-2}}{2N(1-2N)}%
(1-2N)_{n}(2N)_{n},~~n=2,3,...~~~~~~
\end{eqnarray}%
with $b_{1}=(N-2N^{2})\zeta ^{2}$ and $b_{n}=[n(n-1)+2N-4N^{2}]\zeta ^{2}$
for $n=2,3,...$ Due to the fact that $\mathcal{H}_{N}$ is non-Hermitian when 
$N\neq 1/4$ these norms are not positive definite even for non-vanishing
polynomials. However, for $N=1/4$ the expressions reduce to the positive
definite norms%
\begin{equation}
N_{n}^{P}=\frac{\zeta ^{2n}}{2\pi }\Gamma ^{2}\left( \frac{1}{2}+n\right)
,\qquad \text{and\qquad }N_{n}^{Q}=4\frac{\zeta ^{2n-2}}{\pi }\Gamma
^{2}\left( \frac{1}{2}+n\right) .
\end{equation}%
So far we did not require the explicit expressions for the measure, but as
argued in \cite{Krajewska} the concrete formulae for $\omega (E)$ may be
computed from 
\begin{equation}
\omega (E)=\sum\limits_{k=1}^{L}\omega _{k}\delta (E-E_{k}),
\end{equation}%
where the energies $E_{k}$ are the $L$ roots of the polynomial $\Phi $ and
the $L$ constants $\omega _{k}$ are determined by the $L$ equations%
\begin{equation}
\sum\limits_{k=1}^{L}\omega _{k}\Phi _{n}(E_{k})=\delta _{n0}\text{,}\qquad 
\text{for }n\in \mathbb{N}_{0}.  \label{om}
\end{equation}%
When $\Phi =P$ we have $L=2N$ and for $\Phi =Q$ the upper limit is $L=2N-1$.

As examples, we solve these equations for the even solutions with $N=1$ to%
\begin{equation}
\omega _{\pm }^{c}=\frac{1}{2}\pm \frac{1}{\sqrt{4-\zeta ^{2}}},
\end{equation}%
such that%
\begin{equation}
N_{1}^{P}=\mathcal{L}(P_{1}^{2})=\omega _{+}^{c}\left( E_{2}^{c,-}\right)
^{2}+\omega _{-}^{c}\left( E_{2}^{c,+}\right) ^{2}=b_{1}=-\zeta ^{2}.
\end{equation}%
Similarly we find for the odd solutions with $N=3/2$%
\begin{equation}
\omega _{\pm }^{s}=\frac{1}{2}\pm \frac{3}{\sqrt{36-\zeta ^{2}}},
\end{equation}%
such that%
\begin{equation}
N_{2}^{Q}=\mathcal{L}(Q_{2}^{2})=\omega _{+}^{s}\left( E_{3}^{s,-}\right)
^{2}+\omega _{-}^{s}\left( E_{3}^{s,+}\right) ^{2}=b_{2}=-4\zeta ^{2}.
\end{equation}

We also compute the moment functionals defined in \cite{Favard,Finkel} as%
\begin{equation}
\mu _{n}:=\mathcal{L}(E^{n})=\sum\limits_{k=1}^{L}\omega
_{k}E_{k}^{n}=\sum\limits_{k=0}^{n-1}\nu _{k}^{(n)}\mu _{k},  \label{mu}
\end{equation}%
Once again also these quantities can be obtained in two alternative ways,
that is either from the computation of the integrals or directly from the
original polynomials $P_{n}$ and $Q_{n}$ without the knowledge of the
constants $\omega _{k}$. In the last equation the coefficients $\nu
_{k}^{(n)}$ are defined through the expansion $P_{n}(E)=2^{n-1}E^{n}-\sum%
\nolimits_{k=0}^{n-1}\nu _{k}^{(n)}E^{k}$ \ and $Q_{n}(E)=2^{n-1}E^{n-1}-%
\sum\nolimits_{k=0}^{n-2}\nu _{k}^{(n)}E^{k}$ for the even and odd
solutions, respectively. For the above even solutions with $N=1$ we obtain%
\begin{eqnarray}
\mu _{0}^{P} &=&1, \\
\mu _{1}^{P} &=&0=\nu _{0}^{(1)}\mu _{0}^{P}, \\
\mu _{2}^{P} &=&-\zeta ^{2}=\frac{1}{2}\left( \nu _{0}^{(2)}\mu _{0}^{P}+\nu
_{1}^{(2)}\mu _{1}^{P}\right) =-\frac{1}{2}2\zeta ^{2}, \\
\mu _{3}^{P} &=&-4\zeta ^{2}=\frac{1}{4}\left( 64\zeta ^{2}-80\zeta
^{2}\right) , \\
\mu _{4}^{P} &=&-16\zeta ^{2}+\zeta ^{4}=\frac{1}{8}\left( -4608\zeta
^{2}+8\zeta ^{4}+6272\zeta ^{2}-448\times 4\zeta ^{2}\right) , \\
\mu _{5}^{P} &=&-64\zeta ^{2}+8\zeta ^{4},
\end{eqnarray}%
and similarly for the odd solutions with $N=3/2$ we compute%
\begin{eqnarray}
\mu _{0}^{Q} &=&1, \\
\mu _{1}^{Q} &=&4, \\
\mu _{2}^{Q} &=&16-\zeta ^{2}, \\
\mu _{3}^{Q} &=&64-24\zeta ^{2}, \\
\mu _{4}^{Q} &=&256-432\zeta ^{2}+\zeta ^{4}, \\
\mu _{5}^{Q} &=&1024-7168\zeta ^{2}+44\zeta ^{4}.
\end{eqnarray}%
\qquad

Thus we have demonstrated here that the model $\mathcal{H}_{N}$, as defined
in (\ref{newH}), does indeed constitute a quasi-exactly solvable model of $%
E_{2}$-type.

\newif\ifabfull\abfulltrue


\end{document}